\documentclass[letterpaper, 10 pt, conference]{ieeeconf}  % Comment this 

\IEEEoverridecommandlockouts                              
\usepackage[pdftex]{graphicx}
\usepackage{amsmath}
\usepackage{subfigure}
\usepackage{tabularx}
\usepackage{cite}
\usepackage{color}
\usepackage{stfloats}
\usepackage[bookmarks=true,colorlinks,linkcolor=black,citecolor=black,urlcolor=black]{hyperref}

\graphicspath{{Graphics/}}

\title{\LARGE \bf Modeling and Control of a Cable-Driven Series Elastic Actuator}

\author{Wulin Zou, Ningbo Yu \\
	\thanks{Mr. Wulin Zou and Assoc. Prof. Dr. Ningbo Yu are with the Institute of Robotics and Automatic Information Systems, Nankai University, and Tianjin Key Laboratory of Intelligent Robotics, Nankai University, Haihe Education Park, Tianjin 300350, China.}%
	\thanks{E-mail: wlzou@mail.nankai.edu.cn, nyu@nankai.edu.cn}%
}

\begin{document}

\maketitle
\thispagestyle{empty}
\pagestyle{empty}

\begin{abstract}
	
Series elastic actuators (SEA) are playing an increasingly important role in the fields of physical human-robot interaction. This paper focuses on the modeling and control of a cable-driven SEA. First, the scheme of the cable-driven SEA has been proposed, and a velocity controlled DC motor has been used as its power source. Based on this, the model of the cable-driven SEA has been built up. Further, a two degrees of freedom (2-DOF) control approach has been employed to control the output torque. Simulation results have shown that the 2-DOF method has achieved better robust performance than the PD method.

\end{abstract}

\section{Introduction}

Series elastic actuator was first recommended by Pratt and Williamson for its benefits of greater shock tolerance, accurate and stable force output and the ability of energy storage \cite{Pratt1995}. Nowadays, SEAs are widely applied in various physical human-robot interaction applications, e.g., the SEA for walking robots \cite{Robinson1999}, the compact rotary SEA for human assistive robots \cite{Kong2012}, the compact compliant actuator for rehabilitation robots \cite{Yu2013}, etc.

Cable actuation has attracted intensive research for its advantages of low inertia, flexible installation, remote actuation, etc. Cables with low weight to length ratio can change the force direction intentionally and easily, enable power transmission to a remote distance with less energy loss and space occupation \cite{Chapuis2006,Caverly2015}, and allow detachment of the actuation motor from the robot frame \cite{Veneman2006}. Besides, a cable actuated system can be a safe solution due to its unidirectional force constraint and property of breaking when the tension exceeding the threshold. Many cable actuated systems have been applied for physical human-robot interaction, such as the LOwer extremity Powered ExoSkeleton (LOPES) \cite{Veneman2006}, the Universal Haptic Drive (UHD) for upper extremity rehabilitation \cite{Oblak2010}, the MR-compatible wrist robot \cite{Sergi2015}, etc.

The force/torque controller of a series elastic actuator is designed to generate a spring deflection force/torque to follow a given command trajectory. Various control approaches have been developed for different SEAs to achieve high force/torque tracking performance. A Proportional-Integral-Differential (PID) control method was introduced in \cite{Robinson2000} to illustrate the concept and performance of a SEA explicitly. A lot of researchers have made important contributions to the development of SEA based on PID control \cite{Wyeth2006,Vallery2007,Oblak2010,Accoto2013,Sergi2015}. Disturbance observer (DOB) based control methods have been also adopted in \cite{Kong2009,Yoo2015,Lu2015} to enhance robustness. Wyeth proposed a cascaded torque control method with an inner velocity loop to overcome problems of non-linearities \cite{Wyeth2006}.

In this paper, the torque controller is designed using the 2-DOF method to track torque reference, eliminate disturbance and noise. Horowitz revealed that the 1-DOF configuration is unable to cope with the two problems of achieving a desired tracking response and attaining a good disturbance/noise rejection at the same time, but the 2-DOF controller could achieve these two goals simultaneously \cite{Horowitz1963}. The reason is that the 2-DOF controller provides an independent way to design tracking response and optimize disturbance/noise rejection. Systematic introduction and design procedures about the 2-DOF control method can be found in \cite{Qiu2010,Huang2015}. This paper is also an extensive work of \cite{ZouWulin2016CCC,ZouWulin2016IROS,ZouWulin2016JAS}.

This paper is organized as follows. Section~\ref{section2} describes the concept and model of the velocity sourced cable-driven SEA. Details for the torque controller design are given in Section~\ref{section3}. Simulations and results are presented in Section~\ref{section4}. Section~\ref{section5} concludes the paper.

\section{The Cable-Driven SEA}\label{section2}

\subsection{The Model of the Cable-Driven SEA}

\begin{figure}[!htbp]
	\centering
	\includegraphics[width=0.8\columnwidth]{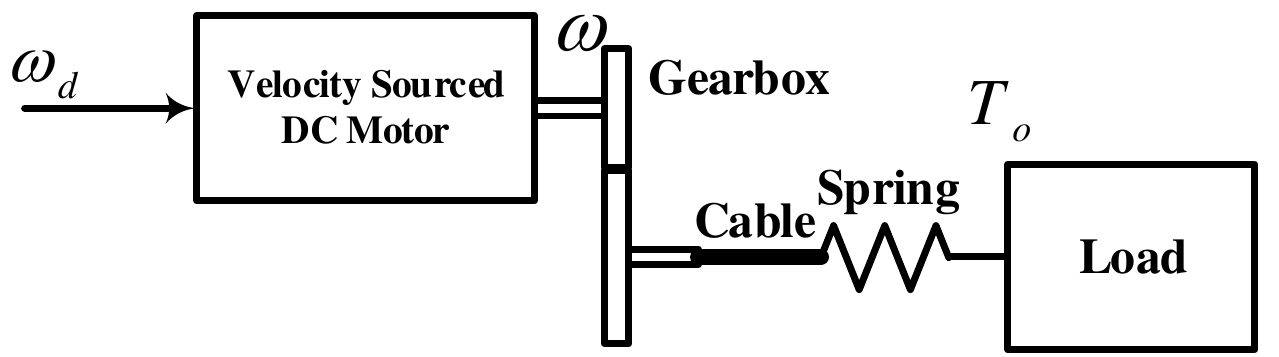}
	\caption{The velocity sourced cable-driven SEA.}\label{fig_Cable_SEA}
\end{figure}

The principle of  a cable-driven SEA is shown in Fig.~\ref{fig_Cable_SEA}. A cable-spring series structure is introduced between the motor and the load. The power source of the actuator is supplied by a velocity controlled DC motor.

Neglecting the deflection of the cable, the velocity of the cable is given as 
\begin{equation}\label{eq11}
{\omega _m} = \frac{\omega }{{{K_g}}}.
\end{equation}
Where, \(K_g\) is the gearbox ratio and $\omega$ is the output velocity of the DC motor. Then, the displacement of the cable is
\begin{equation}\label{eq12}
{\theta _m} = \frac{{{\omega _m}}}{s}.
\end{equation}

If the displacement of the load is $\theta$\(_l\), the spring deflection $\theta$\(_s\) can be derived as
\begin{equation}\label{eq13}
{\theta _s} = {\theta _m} - {\theta _l}.
\end{equation}
The torque \(T_o\) applied to the load is due to the deflection of the spring, such that
\begin{equation}\label{eq14}
{T_o} = {K_s}{\theta _s}.
\end{equation}
If the inertia and damping of the spring are considered, it becomes
\begin{equation}\label{eq15}
{T_o} = ({M_s}{s^2} + {C_s}s + {K_s})({\theta _m-\theta _l}).
\end{equation}

Let $\theta$\(_l=0\) and combine (\ref{eq11}), (\ref{eq12}) and (\ref{eq15}), there is
\begin{equation}\label{eq16}
\frac{{{T_o}(s)}}{{\omega (s)}} = \frac{{{M_s}{s^2} + {C_s}s + {K_s}}}{{{K_g}s}}.
\end{equation}

\subsection{Velocity Sourced DC Motor}

To design a well performed SEA, a DC motor is used as the velocity source of the actuator. The reason that this idea is adopted is that velocity control is easier and more straightforward than current control. What's more, velocity control can overcome some undesirable effects caused by motor internal disturbance.

\begin{figure}[!htbp]
	\centering
	\includegraphics[width=\columnwidth]{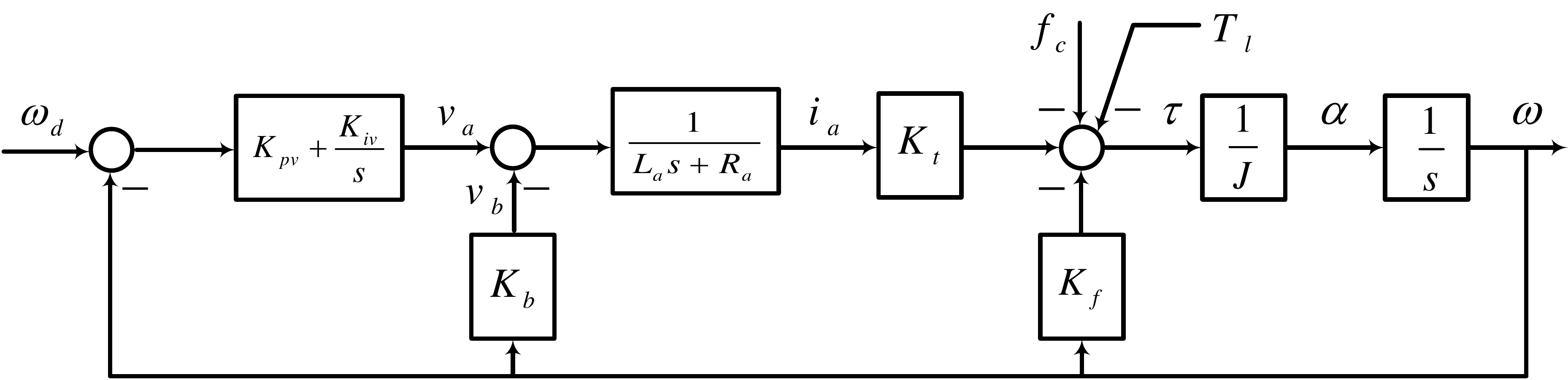}
	\caption{The diagram of the velocity sourced DC motor.}\label{fig1}
\end{figure}

The schematic diagram of a velocity sourced DC motor is shown in Fig.~\ref{fig1}. The velocity feedback regulated by a well tuned PI controller forms a stable closed loop so that the DC Motor system can track the reference signal quickly and accurately. 

When the current \(i_a\) flows through the motor coil, there is
\begin{equation}\label{eq1}
{L_a}\frac{{d{i_a}}}{{dt}} + {i_a}{R_a} + {v_b} = {v_a}.
\end{equation}
Where \(v_a\) is the applied voltage across the motor terminals, \(R_a\) is the resistance of the motor winding, \(L_a\) is the inductance of the motor coil, and \(v_b\) is the back emf voltage that is linearly proportional to the angular velocity $\omega$ of the motor, such that \(v_b = K_b\omega\) .

The equation of the shaft rotating about a fixed axis is
\begin{equation}\label{eq2}
{K_t}{i_a} - {K_f}\omega - {f_c} - {T_l} = J\alpha.
\end{equation}
Where $\alpha$ is the angular acceleration such that $\alpha$ \(=\) $\omega$\(s\). The magnetic torque is linearly proportional (\(K_t\)) to the current \(i_a\) flowing through the motor coil. The viscous friction torque is in the opposite direction of the motion and is linearly proportional (\(K_f\)) to the angular velocity $\omega$. \(f_c\) is the Coulomb friction and \(T_l\) is the load torque.

Substitute (\ref{eq1}) into (\ref{eq2}) and convert it to frequency domain equation, $\omega$(\(s\)) can be obtained as
\begin{equation}\label{eq3}
\omega (s) = \frac{{{v_a}(s){K_t} - ({L_a}s + {R_a})({f_c}(s) + {T_l}(s))}}{{J{L_a}{s^2} + (J{R_a} + {K_f}{L_a})s + {K_f}{R_a} + {K_b}{K_t}}}.
\end{equation}

\subsection{Design of the Velocity Controller}

In order to control the DC motor to be an effective velocity source, a PI controller is designed there. As shown in Fig.~\ref{fig1}, the velocity controller has the form of
\begin{equation}\label{eq4}
{C_v}(s) = {K_{pv}} + \frac{{{K_{iv}}}}{s}.
\end{equation}

The velocity controller is tuned in the case of no load attached and neglecting the Coulomb friction, that is to say, the load torque \(T_l\) and the Coulomb friction \(f_c\) are set as zero. To create a system that will be of type 1, change (\ref{eq3}) and (\ref{eq4}) to the following form:
\begin{equation}\label{eq5}
{G_v}(s) = \frac{{\omega (s)}}{{{v_a}(s)}} = \frac{{\frac{{{K_t}}}{{J{L_a}}}}}{{(s + {p_1})(s + {p_2})}},
\end{equation}

\begin{equation}\label{eq6}
{C_v}(s) = \frac{{{K_{pv}}(s + \frac{{{K_{iv}}}}{{{K_{pv}}}})}}{s}.
\end{equation}

Where \(p_2>p_1>0\) and both can be obtained from (\ref{eq3}) and (\ref{eq5}). Let
\begin{equation}\label{eq7}
\frac{{{K_{iv}}}}{{{K_{pv}}}}=p_1,
\end{equation}
then, the open loop transfer function becomes
\begin{equation}\label{eq8}
\begin{aligned}
{H_{open}} & = {C_v}(s){G_v}(s)\\
& = \frac{{{K_{pv}}(s + \frac{{{K_{iv}}}}{{{K_{pv}}}})}}{s} \cdot \frac{{\frac{{{K_t}}}{{J{L_a}}}}}{{(s + {p_1})(s + {p_2})}}\\
& = \frac{{\frac{{{K_{pv}}{K_t}}}{{J{L_a}}}}}{{s(s + {p_2})}}.
\end{aligned}
\end{equation}

The closed loop transfer function will become a typical two order system, and has the form of
\begin{equation}\label{eq9}
{H_{closed}} = \frac{{\omega _n^2}}{{{s^2} + 2\xi {\omega _n}s + \omega _n^2}}.
\end{equation}

According to (\ref{eq7}), (\ref{eq8}) and (\ref{eq9}), there are
\begin{equation}\label{eq10}
\left\{ {\begin{array}{*{20}{c}}
	{{K_{pv}} = \displaystyle \frac{{\omega _n^2J{L_a}}}{{{K_t}}}} \vspace{2mm}\\
	{{K_{iv}} = {p_1}{K_{pv}}} \vspace{2mm}\\
	{2\xi {\omega _n} = {p_2}}
	\end{array}} \right..
\end{equation}

Given the desired performance of the closed loop system, the natural frequency $\omega$\(_n\) and damping ratio $\xi$ can be estimated. Then, the controller parameters can be obtained from (\ref{eq10}). Further, the transfer function from the desired velocity input $\omega$\(_d(s)\) to the actuator output torque \(T_o(s)\) can be obtained by	$P(s)=\displaystyle \frac{T_o(s)}{\omega_d(s)}$.

\section{Torque Controller Design}\label{section3}

\subsection{Controller with Optimal Transient}

In this subsection, a possible way to design the best stabilizing controller and measure the quality will be introduced. Assume that \(P(s)\) is strictly proper and denoted as the form of
\begin{equation}\label{eq17}
P(s) = \frac{{b(s)}}{{a(s)}} = \frac{{{b_1}{s^{n - 1}} +  \cdots  + {b_n}}}{{{a_0}{s^n} + {a_1}{s^{n - 1}} +  \cdots  + {a_n}}}.
\end{equation}
Where \(a(s)\) and \(b(s)\) are coprime and \(a_0 \ne 0\).

\begin{figure}[!htbp]
	\centering
	\includegraphics[width=0.55\columnwidth]{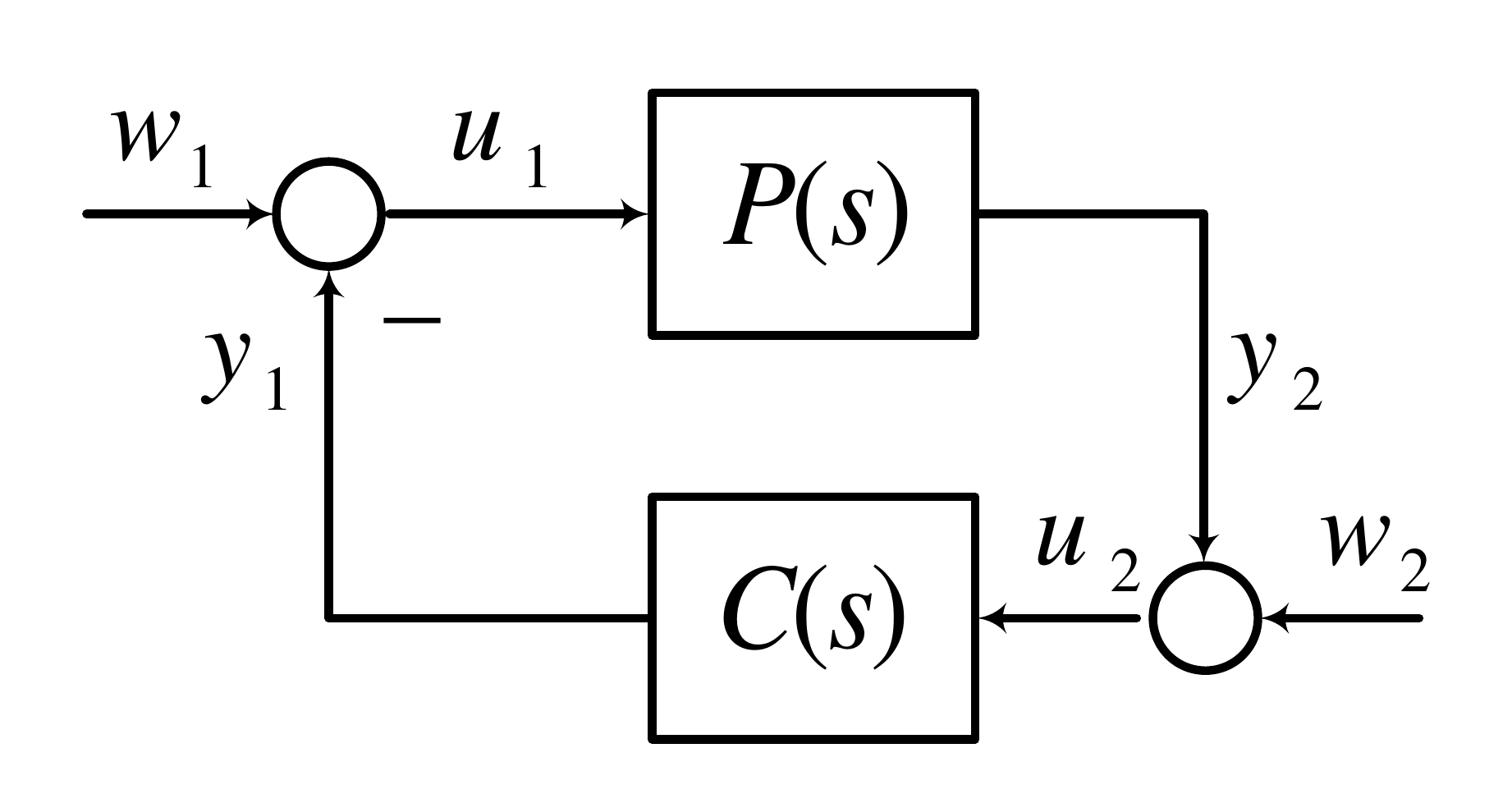}
	\caption{Feedback system for stabilization.}\label{fig4}
\end{figure}

Consider the system shown in Fig.~\ref{fig4}, if the closed loop system is internally stable and the external inputs \(w_i(t),i=1,2,\) are all impulse signals, all the internal signals \(u_j(t),y_j(t),j=1,2,\) will eventually settle to zero when time goes to infinity. So, the RMS value of \(y_i(t)\,i=1,2,\) when \(w_i(t),i=1,2,\) are unit impulses can be used to measure the quality of the performance as follow:
\begin{equation}\label{eq18}
\begin{aligned}
J = &(||{y_1}(t)||_2^2 + ||{y_2}(t)||_2^2){|_{\scriptstyle{w_1}(t) = \delta (t)\hfill\atop\scriptstyle{w_2}(t) = 0\hfill}}\\
&+ (||{y_1}(t)||_2^2 + ||{y_2}(t)||_2^2){|_{\scriptstyle{w_1}(t) = 0\hfill\atop\scriptstyle{w_2}(t) = \delta (t)\hfill}}.
\end{aligned}
\end{equation}

For different stabilizing controllers, their RMS value \(J\) are different. The minimum value of \(J\) denoted by \(J^*\) when \(C(s)\) is chosen among all stabilizing controllers means that the best performance is achieved.

The transfer function from (\(w_1,\)\(w_2\)) to (\(y_1,\)\(y_2\)) is:
\begin{equation}\label{eq19}
\left[ \begin{array}{l}
\begin{array}{*{20}{c}}
\displaystyle{\frac{{P(s)C(s)}}{{1 + P(s)C(s)}}}&\displaystyle{\frac{{C(s)}}{{1 + P(s)C(s)}}}\vspace{2mm}\\
\displaystyle{\frac{{P(s)}}{{1 + P(s)C(s)}}}&\displaystyle{\frac{{ - P(s)C(s)}}{{1 + P(s)C(s)}}}
\end{array}
\end{array} \right].
\end{equation}

The next problem to be solved is how to find the most optimal stabilizing controller. Firstly, there must exist a stable polynomial \(d(s)\) called spectral factor of \(a( - s)a(s) + b( - s)b(s)\) such that:
\[a( - s)a(s) + b( - s)b(s) = d({\rm{ - }}s)d(s).\]

Now let \(C(s) = \displaystyle \frac{{q(s)}}{{p(s)}}\) be the unique n-th order strictly proper pole placement controller such that:
\[a(s)p(s) + b(s)q(s) = {d^2}(s).\]

Then, this controller \(C(s)\) is the most optimal controller which minimizes \(J\) and gives
\begin{equation}\label{eq21}
\begin{aligned}
{J^*} =& ||\frac{{d(s) - a(s)}}{{d(s)}}||_2^2 + ||\frac{{b(s)}}{{d(s)}}||_2^2\\
&+ ||\frac{{d(s) - p(s)}}{{d(s)}}||_2^2 + ||\frac{{q(s)}}{{d(s)}}||_2^2.
\end{aligned}
\end{equation}

If \(w_i,i=1,2,\) are external disturbances or noises, a stabilizing controller with optimal transient can eliminate them in a short time.

\subsection{Stabilizing 2-DOF Controller}

\begin{figure}[!htbp]
	\centering
	\includegraphics[width=0.6\columnwidth]{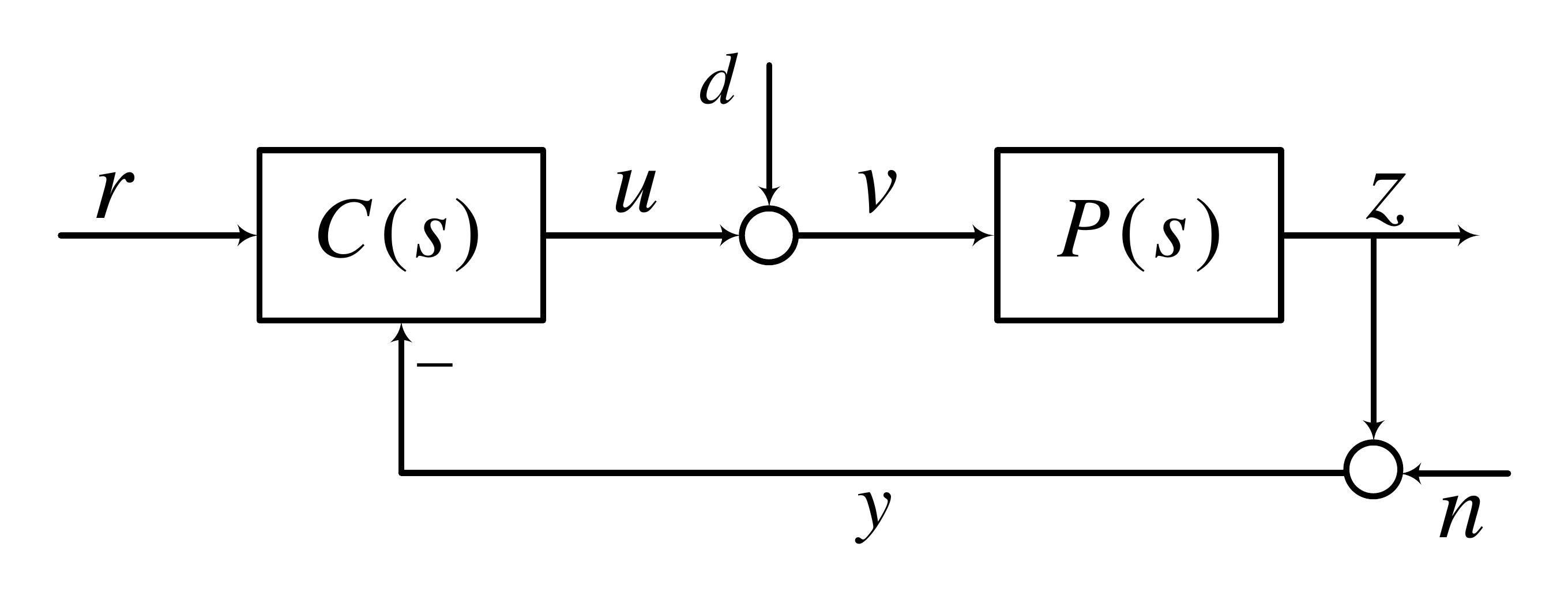}
	\caption{The 2-DOF control configuration.}\label{fig3}
\end{figure}

A closed loop system using 2-DOF controller to stabilize the plant \(P(s)\) is shown in Fig.~\ref{fig3}. The output \(z\) can track the reference input signal \(r\) in a satisfactory way, also, the disturbance \(d\) and the noise \(n\) can be eliminated close to zero in a short time.

\begin{figure}[!htbp]
	\centering
	\includegraphics[width=0.8\columnwidth]{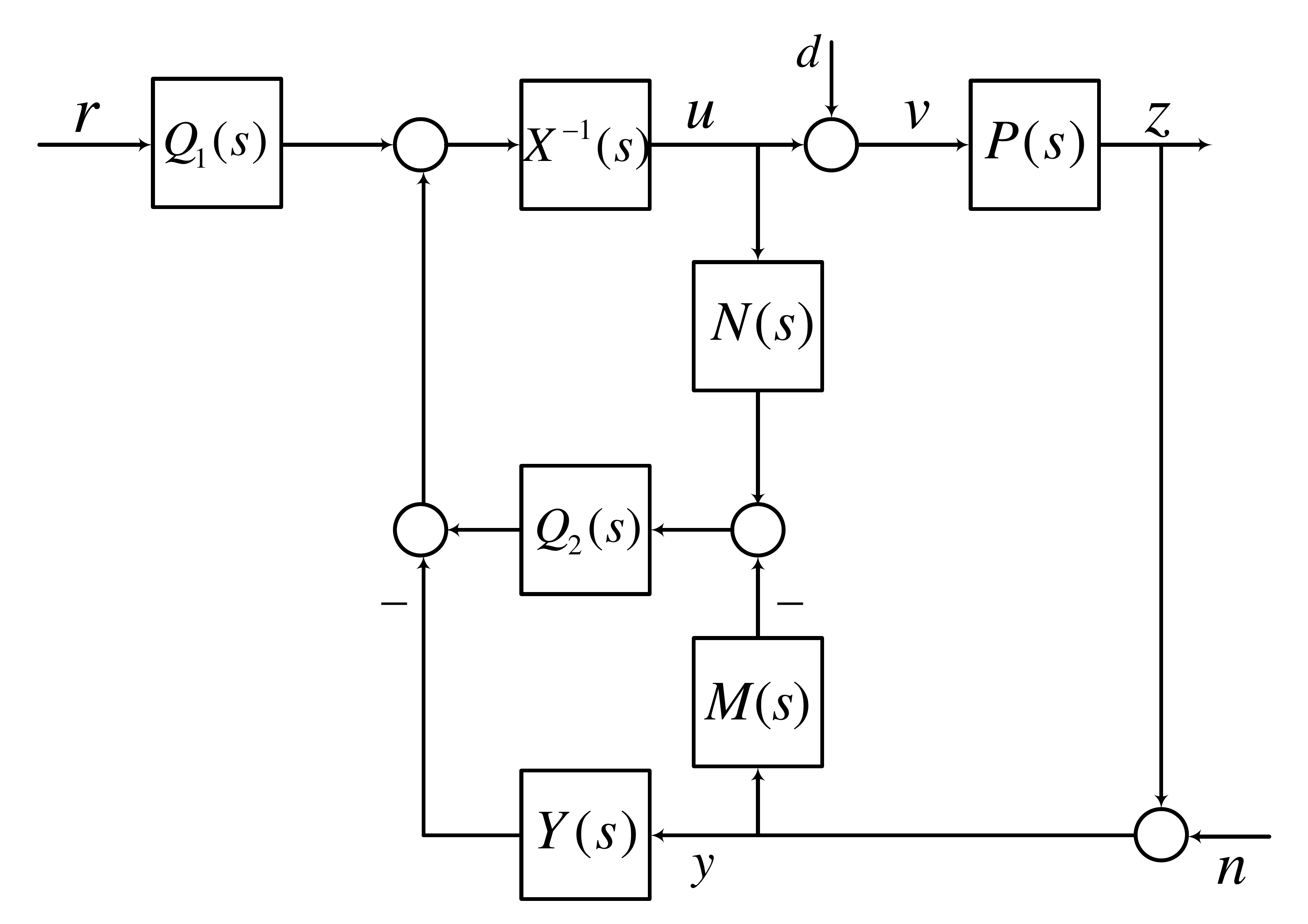}\\
	\caption{Stabilizing 2-DOF control structure.}\label{fig5}
\end{figure}

The stabilizing 2-DOF controller based on the controller with optimal transient is shown in Fig.~\ref{fig5}. Let \(C_0(s)\) be a stabilizing controller with optimal transient: \(C_0(s) = \displaystyle \frac{{q(s)}}{{p(s)}}\). Let \(M(s),N(s),X(s),Y(s)\) be:
\begin{equation}\label{eq22}
\begin{aligned}
&M(s) = \frac{{a(s)}}{{d(s)}},N(s) = \frac{{b(s)}}{{d(s)}},\\
&X(s) = \frac{{p(s)}}{{d(s)}},Y(s) = \frac{{q(s)}}{{d(s)}}.
\end{aligned}
\end{equation}

Then \(M(s),N(s),X(s),Y(s)\) are all stable transfer functions satisfying:
\[P(s) = \frac{{N(s)}}{{M(s)}},{C_0}(s) = \frac{{Y(s)}}{{X(s)}},\]
\[M(s)X(s) + N(s)Y(s) = 1.\]

Denote the set of all the 2-DOF controllers which give stable closed loop systems by \(\Omega (P)\):
\begin{equation}\label{eq23}
\Omega (P) = \{ C(s) = \left[ {\begin{array}{*{20}{c}}
	{\frac{{{Q_1}(s)}}{{X(s) - N(s){Q_2}(s)}}}&{\frac{{Y(s) + M(s){Q_2}(s)}}{{X(s) - N(s){Q_2}(s)}}}
	\end{array}} \right]\}.
\end{equation}
Where, \(Q_1(s),Q_2(s)\) are arbitrary stable transfer functions. 

Every stabilizing 2-DOF controller has the form shown in Fig.~\ref{fig5}. For a fixed plant \(P(s)\), if we plug a stabilizing 2-DOF controller of this form to a closed loop transfer function, then the closed loop transfer function becomes a function of \(Q_1(s)\) and \(Q_2(s)\). The problem becomes choosing good \(Q_1(s)\) and \(Q_2(s)\) to meet the design specifications, as long as they are stable. 

The four transfer functions from \(r\) to the internal variables \(u,v,y,z\) depend only on \(Q_1(s)\):
\begin{equation}\label{eq24}
\left[ {\begin{array}{*{20}{c}}
	\displaystyle{\frac{{{C_1}(s)}}{{1 + P(s){C_2}(s)}}}\vspace{2mm}\\
	\begin{array}{l}
	\displaystyle\frac{{{C_1}(s)}}{{1 + P(s){C_2}(s)}}\vspace{2mm}\\
	\displaystyle\frac{{P(s){C_1}(s)}}{{1 + P(s){C_2}(s)}}\vspace{2mm}\\
	\displaystyle\frac{{P(s){C_1}(s)}}{{1 + P(s){C_2}(s)}}
	\end{array}
	\end{array}} \right]{\rm{ = }}\left[ \begin{array}{l}
M(s)\\
M(s)\\
N(s)\\
N(s)
\end{array} \right]{Q_1}(s).
\end{equation}

The eight transfer functions from \(d,n\) to \(u,v,y,z\) depend only on \(Q_2(s)\):
\begin{equation}\label{eq25}
\begin{array}{l}
\left[ \begin{array}{l}
\begin{array}{*{20}{c}}
\displaystyle{\frac{{ - P(s){C_2}(s)}}{{1 + P(s){C_2}(s)}}}&\displaystyle{\frac{{ - {C_2}(s)}}{{1 + P(s){C_2}(s)}}}
\end{array}\vspace{2mm}\\
\begin{array}{*{20}{c}}
\displaystyle{\frac{1}{{1 + P(s){C_2}(s)}}}&\displaystyle{\frac{{ - {C_2}(s)}}{{1 + P(s){C_2}(s)}}}
\end{array}\vspace{2mm}\\
\begin{array}{*{20}{c}}
\displaystyle{\frac{{P(s)}}{{1 + P(s){C_2}(s)}}}&\displaystyle{\frac{1}{{1 + P(s){C_2}(s)}}}
\end{array}\vspace{2mm}\\
\begin{array}{*{20}{c}}
\displaystyle{\frac{{P(s)}}{{1 + P(s){C_2}(s)}}}&\displaystyle{\frac{{ - P(s){C_2}(s)}}{{1 + P(s){C_2}(s)}}}
\end{array}
\end{array} \right] = \\
\left[ {\begin{array}{*{20}{c}}
	{ - N(s)Y(s)}&{ - M(s)Y(s)}\\
	\begin{array}{l}
	M(s)X(s)\\
	N(s)X(s)\\
	N(s)X(s)
	\end{array}&\begin{array}{l}
	- M(s)Y(s)\\
	M(s)Y(s)\\
	- N(s)X(s)
	\end{array}
	\end{array}} \right]\\
- \left[ \begin{array}{l}
M(s)\\
M(s)\\
N(s)\\
N(s)
\end{array} \right]{Q_2}(s)\left[ {\begin{array}{*{20}{c}}
	{N(s)}&{M(s)}
	\end{array}} \right]
\end{array}.
\end{equation}

This makes choosing \(Q_1(s)\) and choosing \(Q_2(s)\) decoupled and rather convenient.

\subsection{Parameterization of $Q_1(s)$ and $Q_2(s)$}

Since the transfer function from the reference input \(r\) to the output signal \(z\) is \(N(s)Q_1(s)\), let
\[{Q_1}(s) = \frac{{\bar \omega _n^2}}{{N(s)({s^2} + 2\bar \xi {\bar \omega _n}s + \bar \omega _n^2)}}.\]

Then, the transfer function from \(r\) to \(z\)  becomes
\[N(s){Q_1}(s) = \frac{{\bar \omega _n^2}}{{{s^2} + 2\bar \xi {\bar \omega _n}s{\rm{ + }}\bar \omega _n^2}}.\]

Considering that \(Q_2(s)\) is mainly used to filter the disturbance \(d\) and noise \(n\) whose industrial frequencies are assumed to be 50 Hz, let
\[{Q_2}(s) = \frac{1}{{s/(100\pi) + 1}}.\]

\section{Simulations and Results}\label{section4}

The cable-driven SEA parameters are listed in Table \ref{table1}. 
\begin{table}[!htbp]
	\caption{Parameters of the cable-driven SEA.}
	\label{table1}
	\begin{center}
		\begin{tabular}{c|c}
			\hline
			\(J\) & \(6.96 \times {10^{ - 6}}{\rm{ kg}} \cdot {{\rm{m}}^{\rm{2}}}\)\\
			\hline
			\({L_a}\) & \(0.6{\rm{2 mH}}\)\\
			\hline
			\({R_a}\) & \({\rm{2}}{\rm{.07 \Omega }}\)\\
			\hline
			\({K_t}\) & \({\rm{0}}{\rm{.0525 Nm/A}}\)\\
			\hline
			\({K_b}\) & \(0.0525{\rm{ Vs/rad}}\)\\
			\hline
			\({K_f}\) & \(0.00001{\rm{ Nm/(rad/s)}}\)\\
			\hline
			\({K_s}\) & \(138{\rm{ Nm/rad}}\)\\
			\hline
			\({K_g}\) & \(156:1\)\\
			\hline
			\({C_s}\) & \(0.01{\rm{ Nm/(rad/s)}}\)\\
			\hline
			\({J_l}\) & \(0.1{\rm{ kg}} \cdot {{\rm{m}}^2}\)\\
			\hline
			\({M_s}\) & \(0.00001{\rm{ kg}} \cdot {{\rm{m}}^{\rm{2}}}\)\\
			\hline
		\end{tabular}
	\end{center}
\end{table}

\subsection{Simulation Results of the Velocity Control}
The transfer function from \(v_a(s)\) to \(\omega(s)\) is:
\[\frac{{{v_a}(s)}}{{\omega (s)}} = \frac{{{\rm{1}}{\rm{.217e07}}}}{{{s^2} + 3340s + 6.435e05}}.\]

\begin{figure}[!htbp]
	\centering
	\includegraphics[width=0.8\columnwidth]{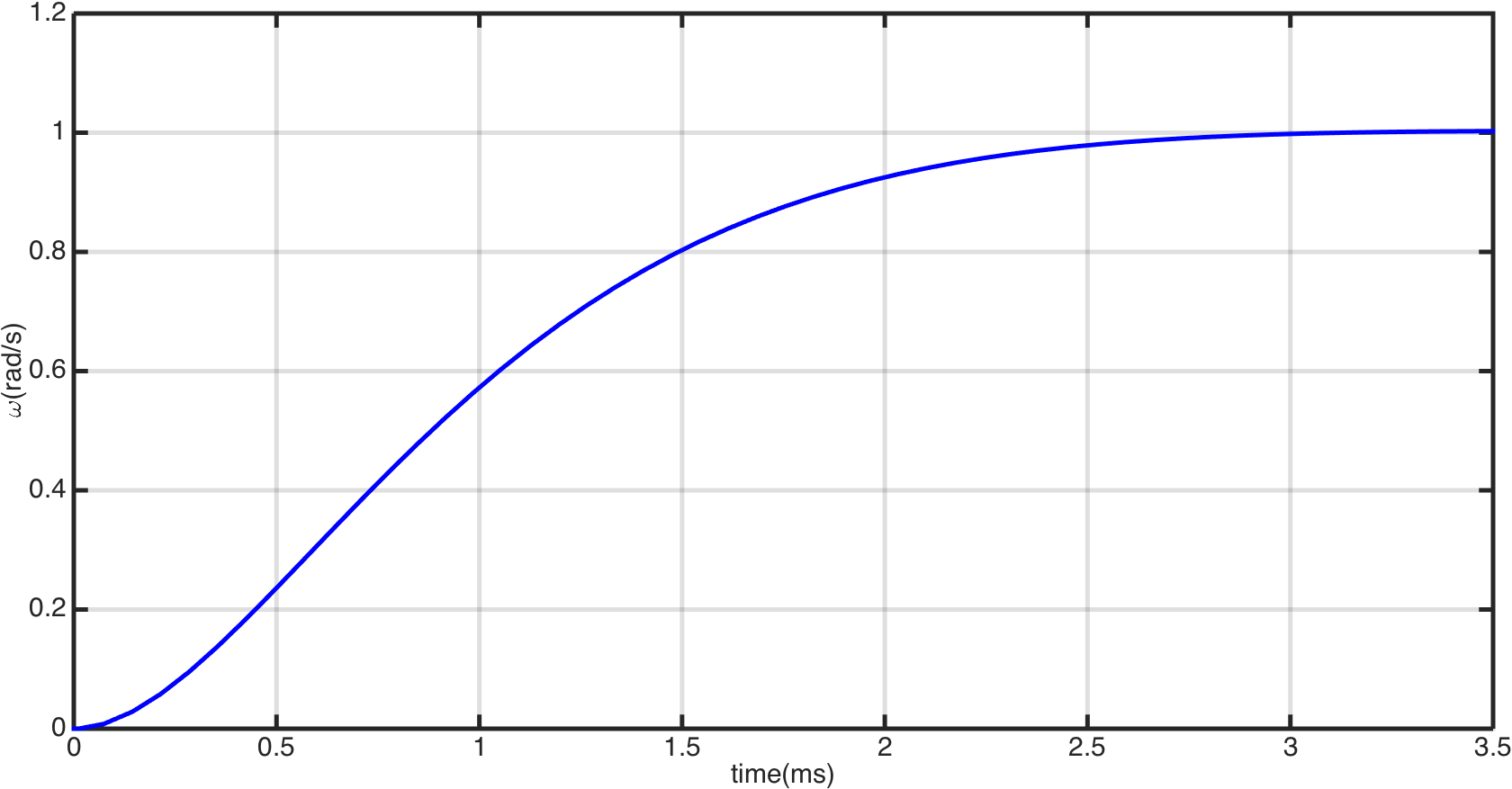}\\
	\caption{Step response of the velocity control}\label{fig6}
\end{figure}

When the damping ratio \(\xi\) is chosen to be 0.88, the coefficients of the proportional-integral controller are \({K_{pv}} = {\rm{0}}{\rm{.26}},{K_{iv}} = {\rm{53}}{\rm{.5}}\). The step response of velocity closed loop system is shown in Fig.~\ref{fig6}. The rise time of the velocity closed loop system is within 0.002s, almost no overshoot, can meet the requirements of a velocity source.

\subsection{Simulation Results of the Torque Control}
The transfer function \(P(s)\) from \(\omega_d(s)\) to \(T_o(s)\) when the load is fixed is:
\[P(s) = \frac{{0.2034{s^3} + 245.1{s^2} + 2.848e06s + 5.761e08}}{{{s^4} + 3340{s^3} + 3.817e06{s^2} + 6.54e08s}}\]

\begin{figure}[!htbp]
	\centering
	\includegraphics[width=0.7\columnwidth]{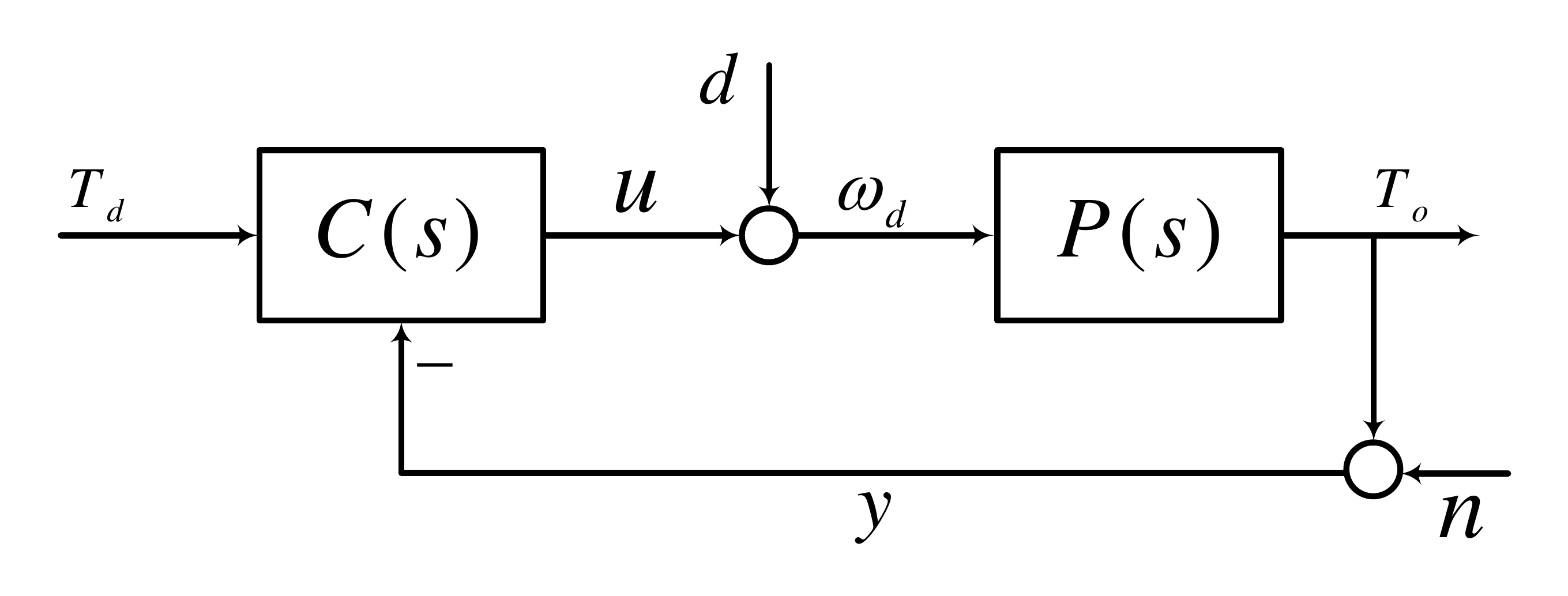}\\
	\caption{Torque control structure for the cable-driven SEA.}\label{fig7}
\end{figure}

The stabilizing 2-DOF controller can be obtained by choosing $\bar \omega _n=451.24$ and $\bar \xi = 0.826$. In the simulation, a PD controller ${C_{PD}}(s) = 490 + 0.1s$ is used for comparison. The torque control structure for the cable-driven SEA is shown in Fig.~\ref{fig7}.

\begin{figure}[!htbp]
	\centering
	\includegraphics[width=0.9\columnwidth]{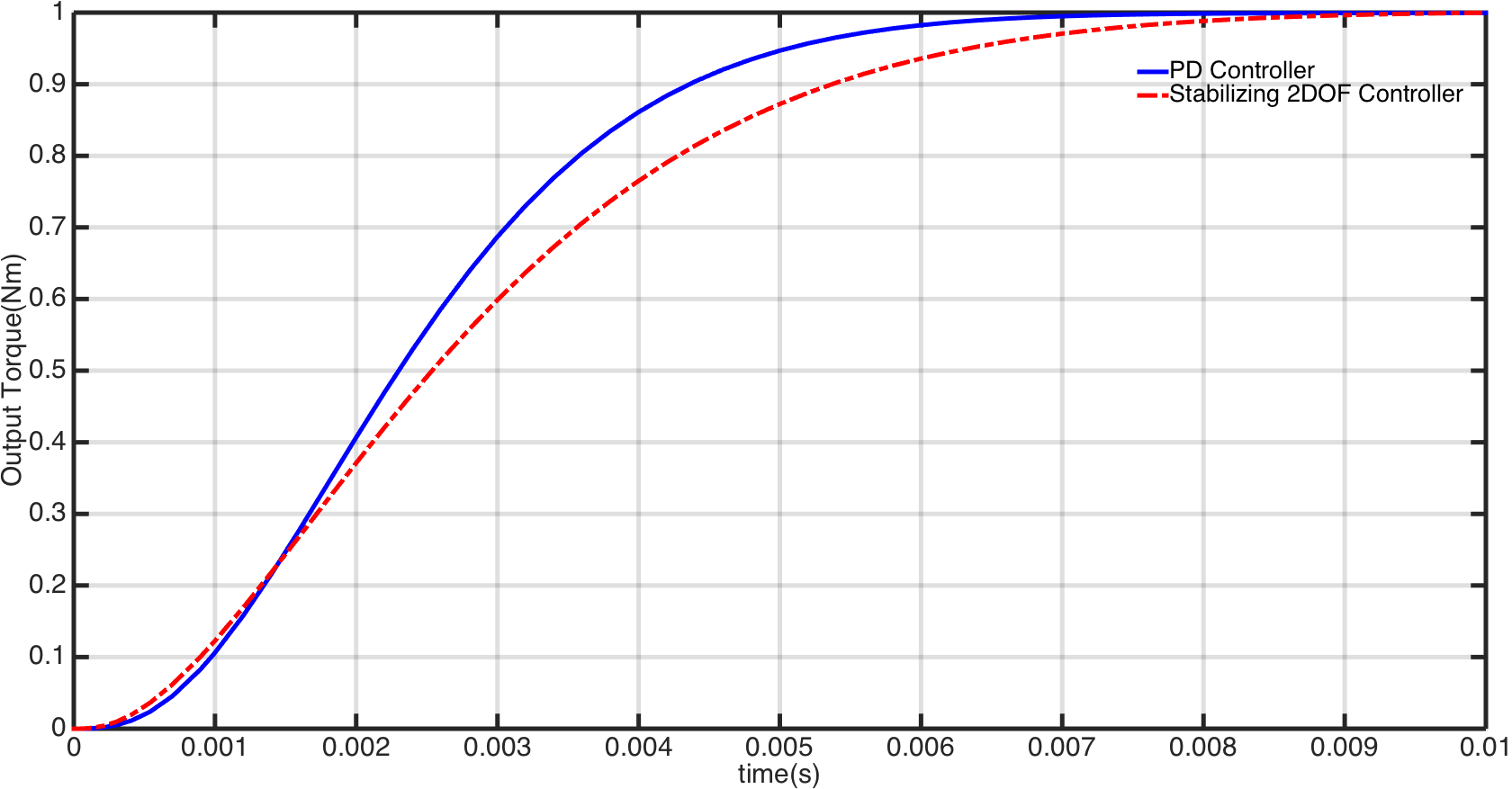}\\
	\caption{Tracking step signal without disturbance and noise}\label{fig8}
\end{figure}

In the first simulation, the disturbance \(d\) and the noise \(n\) are set to zero. Then, the step responses from \(T_d(s)\) to \(T_o(s)\) of the two control method are shown in Fig.~\ref{fig8}. Both the two method tracks the step input quickly and accurately.

\begin{figure}[!htbp]
	\centering
	\includegraphics[width=0.9\columnwidth]{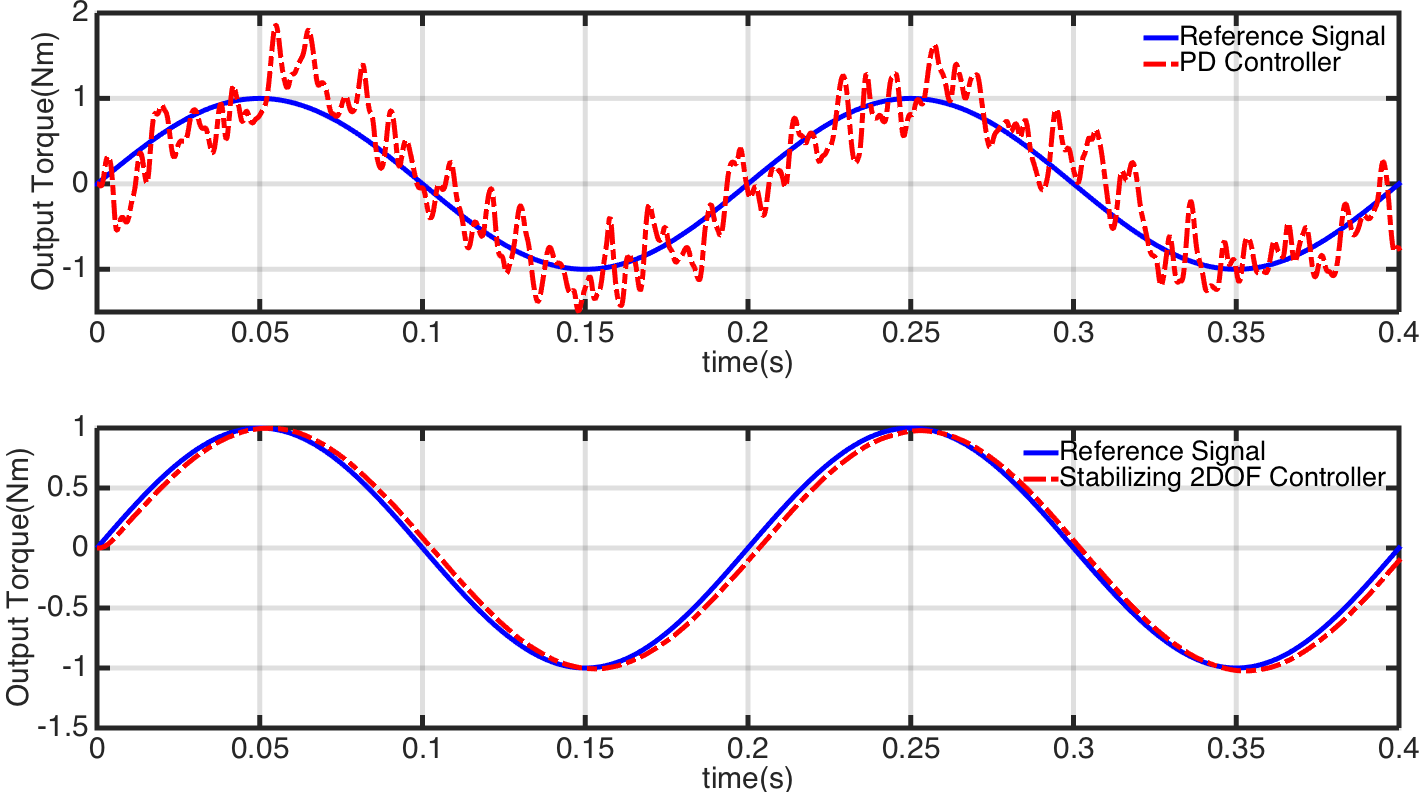}\\
	\caption{Tracking sinusoidal signal with disturbance and noise}\label{fig9}
\end{figure}

In the second simulation, the disturbance \(d\) and the noise \(n\) are set as gaussian white noises with zero mean. The input signal \(T_d\) is set as a sinusoidal signal with frequency of 5 Hz and amplitude of 1 Nm. The tracking results are shown in Fig.~\ref{fig9}. It is clear that the stabilizing 2-DOF controller can enhance the system robustness.

\section{Conclusions}\label{section5}

This paper demonstrates the efficacy of the 2-DOF control approach for the challenging torque control problem of a cable-driven SEA. The 2-DOF torque controller performed better compared with the PD method in the presence of noise and disturbance in the simulations. The torque control performance can be conveniently adjusted by choosing appropriate filter $Q_1(s)$ and $Q_2(s)$.

Further research will be focused on the impedance control of the cable-driven SEA.

\bibliography{reference}
	
\end{document}